# An empirical analysis of the use of alphabetical authorship in scientific publishing

Ludo Waltman

Centre for Science and Technology Studies, Leiden University, The Netherlands
waltmanlr@cwts.leidenuniv.nl

There are different ways in which the authors of a scientific publication can determine the order in which their names are listed. Sometimes author names are simply listed alphabetically. In other cases, authorship order is determined based on the contribution authors have made to a publication. Contribution-based authorship can facilitate proper credit assignment, for instance by giving most credits to the first author. In the case of alphabetical authorship, nothing can be inferred about the relative contribution made by the different authors of a publication.
In this paper, we present an empirical analysis of the use of alphabetical authorship in scientific publishing. Our analysis covers all fields of science. We find that the use of alphabetical authorship is declining over time. In 2011, the authors of less than 4% of all publications intentionally chose to list their names alphabetically. The use of alphabetical authorship is most common in mathematics, economics (including finance), and high energy physics. Also, the use of alphabetical authorship is relatively more common in the case of publications with either a small or a large number of authors.

## 1. Introduction

Scientific publications produced by a single author are becoming more and more uncommon. Of the 1.3 million publications that appeared in 2011 and that have been indexed in the Web of Science database, 89% had more than one author. When a publication has more than one author, the authors need to make a decision on the order in which their names are listed. One way in which authorship order can be determined is simply by listing author names alphabetically. However, many other criteria can be used as well. In particular, authorship order can be determined based on the contribution authors have made to a publication, with the first author being the most significant contributor.

Knowing the way in which authorship order has been determined can be important for proper credit assignment. Suppose we have a highly cited publication with ten authors. If the authors have chosen to list their names alphabetically, the authorship order does not provide us any information on the degree to which each author has contributed to the publication. As a consequence, we have no idea how much each author should be credited for the publication. However, if the authors have chosen to list their names based on the contribution each of them has made, we know that the first author is the most important contributor and, consequently, that the first author deserves more credits than the other authors.[1]

---

[1] The idea of crediting authors based on their position in the author list of a publication was already suggested by Hodge and Greenberg (1981). More recently, there are various papers in which this idea is explored in more detail, often in the context of the *h*-index (e.g., Abbas, 2011; Egghe, Rousseau, & Van Hooydonk, 2000; Galam, 2011; Hagen, 2008, 2010; Hu, Rousseau, & Chen, 2010; Liu & Fang, 2012a, 2012b; Sekercioglu, 2008).

There is a considerable body of literature in which practices for determining authorship order are studied. We refer to Frandsen and Nicolaisen (2010) and Marušić, Bošnjak, and Jerončić (2011) for recent overviews of the literature. Part of the literature focuses on the use of alphabetical authorship. However, although there are various studies in which the use of alphabetical authorship is investigated for specific fields of science (e.g., Frandsen & Nicolaisen, 2010), there are no studies that cover science as a whole. This gap in the literature will be filled in the present paper.

The analysis that we provide in this paper examines the use of alphabetical authorship in all fields of science. Our aim is to determine how often alphabetical authorship is used, how the use of alphabetical authorship increases or decreases over time, and to what extent the use of alphabetical authorship is affected by disciplinary differences. We also study the phenomenon of partial alphabetical authorship, where some of the authors of a publication are listed alphabetically while others are not. An important element in our analysis is the distinction between what we call intentional alphabetical authorship and incidental alphabetical authorship. Intentional alphabetical authorship refers to the situation in which the authors of a publication intentionally choose to list their names alphabetically, while incidental alphabetical authorship refers to the situation in which authors choose to list their names based on a non-alphabetical criterion and in which this criterion incidentally produces an alphabetical authorship order.

The organization of this paper is as follows. In Section 2, the distinction between intentional and incidental alphabetical authorship is discussed in more detail. In Section 3, the empirical analysis is presented. The main conclusions of the analysis are summarized in Section 4.

## 2. Intentional vs. incidental alphabetical authorship

Consider the following situation. Authors Jones and Smith work together on a publication. Jones is the main contributor to the publication. He came up with the original research idea, did most of the empirical work, and also wrote the first draft of the paper. In the field of Jones and Smith, authors usually list their names based on the contribution they have made to a publication, with the first author being the most important contributor. Jones and Smith want to follow this convention, and therefore Jones is listed as the first author and Smith as the second. Incidentally, the order in which Jones and Smith are listed coincides with the alphabetical order. Hence, Jones and Smith are listed alphabetically, even though they chose to be listed based on a non-alphabetical criterion, namely their contribution to the publication. We refer to this situation as incidental alphabetical authorship.

Incidental alphabetical authorship is more likely to occur in the case of a publication with a relatively small number of authors than in the case of a publication with a larger number of authors. To see this, assume that authors list their names based on the contribution they have made to a publication. Also, assume that on average the position of an author name in the alphabet does not correlate with the contribution the author makes to a publication. In other words, it is assumed that on average an author named Anderson does not contribute more or less than an author named Young. Under these assumptions, it is clear that in the case of a publication with two authors there is a 50% probability of incidental alphabetical authorship. In the case of a publication with ten authors, the probability of incidental alphabetical authorship equals $1/10 \times 1/9 \times ... \times 1/1 = 1/10! = 2.8 \times 10^{-7}$, which is a virtually zero probability. Hence, as the number of authors of a publication increases, the probability of incidental alphabetical authorship quickly decreases. This is illustrated in Figure 1.

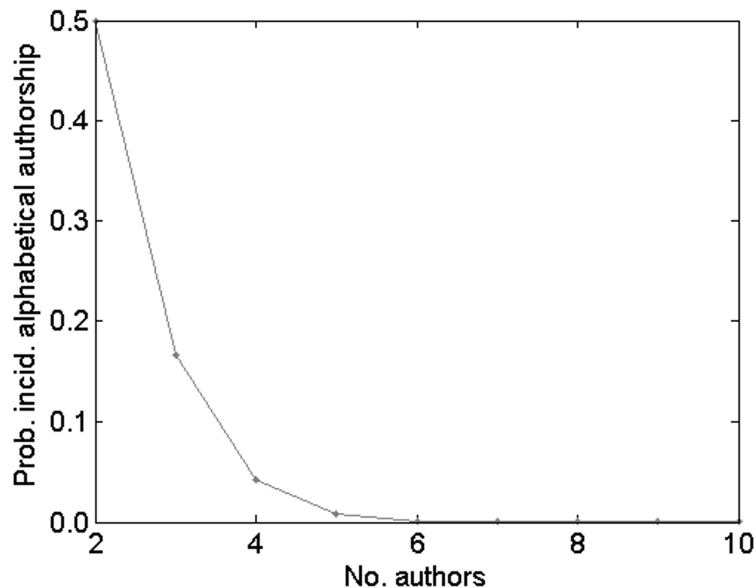

Figure 1. Probability of incidental alphabetical authorship (under the assumptions mentioned in the text) as a function of the number of authors of a publication.

For a proper analysis of the use of alphabetical authorship, we consider it essential to correct for the phenomenon of incidental alphabetical authorship. Without correcting for this phenomenon, it would for instance not be clear how a decrease in the use of alphabetical authorship over time should be interpreted. One interpretation could be that authors of publications less often choose to list their names alphabetically. This interpretation would imply a decrease in intentional alphabetical authorship. However, an alternative interpretation could be that the average number of authors per publication has increased over time and that, as a consequence, there has been a decrease in incidental alphabetical authorship. In order to distinguish between these two interpretations, we need to correct for the phenomenon of incidental alphabetical authorship.

Correcting for the phenomenon of incidental alphabetical authorship requires a model of the way in which authors of publications decide on the order in which their names are listed. The model that we propose assumes that the authors of a publication have two options: They may choose to list their names alphabetically, or they may choose to list their names based on a non-alphabetical criterion.[2] The type of non-alphabetical criterion that is employed is not important, but the model assumes that when the non-alphabetical criterion is used, all possible orderings of author names are equally likely to be observed. Hence, the non-alphabetical criterion must be completely uncorrelated with the alphabetical order of author names.

Let's now formulate our model in more formal terms. Suppose we have a set of $N$ publications, denoted by 1, 2, ..., $N$. Each publication has at least two authors. Let $n_i$ denote the number of authors of publication $i$, and let $p_i$ denote the probability that the authors of publication $i$ intentionally choose to list their names alphabetically. Furthermore, let $a_i = 1$ if the names of the authors of publication $i$ are listed alphabetically, and let $a_i = 0$ if not. Notice that $n_i$ and $a_i$ can be directly observed from the bibliographic data of a publication, while $p_i$ cannot be observed. Our aim is to estimate the average probability that the authors of a publication intentionally choose to list their names alphabetically. In other words, based on $n_1, n_2, ..., n_N$ and $a_1, a_2, ..., a_N$, we wish to estimate

$$\bar{p} = \frac{1}{N} \sum_{i=1}^{N} p_i . \tag{1}$$

---

[2] Our model is similar to a model employed by Van Praag and Van Praag (2008).

Estimating $\bar{p}$ in (1) can be done using the estimator $\hat{p}$ given by

$$\hat{p} = \frac{1}{N} \sum_{i=1}^{N} \hat{p}_i , \tag{2}$$

where

$$\hat{p}_i = \frac{a_i - \frac{1}{n_i!}}{1 - \frac{1}{n_i!}} . \tag{3}$$

We note that $n_i!$ in (3) denotes the factorial of $n_i$, that is, $n_i! = 1 \times 2 \times ... \times n_i$. To show that $\hat{p}$ is an appropriate estimator of $\bar{p}$, we prove in the appendix that the expected value of $\hat{p}$ equals $\bar{p}$. This result indicates that $\hat{p}$ is an unbiased estimator of $\bar{p}$.

There are two comments that we would like to make on the estimator $\hat{p}$:

- If $n_i$ is sufficiently large (e.g., $n_i \geq 5$), $1 / n_i!$ is very close to zero (see Figure 1), which means that $\hat{p}_i$ in (3) is very close to $a_i$. Consequently, in the case of a set of publications that all have a sufficiently large number of authors, $\hat{p}$ in (2) is approximately equal to the proportion publications with alphabetically listed authors. In other words, $\hat{p}$ may deviate from the proportion publications with alphabetically listed authors only if some publications have only a relatively small number of authors. The rationale for this is that the distinction between intentional and incidental alphabetical authorship is relevant only for publications with a relatively small number of authors. In the case of publications with a larger number of authors, incidental alphabetical authorship is highly unlikely to occur, at least under the assumptions that we make in our model.
- Somewhat counterintuitively, it is possible that $\hat{p}$ in (2) is negative. If the number of publications $N$ is relatively small, this may be seen as a kind of small sample effect. In the case of larger $N$, $\hat{p}$ substantially below zero would indicate a model misspecification. It would suggest that the authors of publications intentionally try to avoid being listed alphabetically, which is something that is not anticipated by our model.

## 3. Empirical analysis

We analyze the use of alphabetical authorship in scientific publications in Thomson Reuters' Web of Science database. Our analysis takes into account all publications in the Science Citation Index Expanded, the Social Sciences Citation Index, and the Arts & Humanities Citation Index in the period 1981–2011. Only the document types *article*, *note*, and *review* are considered. There are 24.8 million publications that have one of these document types. Obviously, analyzing the use of alphabetical authorship makes sense only for publications with at least two authors. Our focus therefore is on the 19.6 million multi-author publications in our database.

A question that still remains is what exactly is meant by alphabetical authorship. This may seem obvious, but yet we need some rules for a number of special cases. The rules that we use are as follows:

- The alphabetical order of authors is determined by their last names. If two authors have the same last name, their alphabetical order is determined by their initials.
- Other things equal, a shorter last name precedes a longer one. For instance, if an author has last name 'WILLIAMS' and another author has last name 'WILLIAMSON', the former author precedes the latter one in the alphabetical order.
- If a space, an apostrophe, or a hyphen occurs in an author name, it is ignored. For instance, the last name 'VAN RAAN' is treated as 'VANRAAN'.

Based on the above rules, we have determined for each of our 19.6 million multi-author publications whether the names of the authors are listed alphabetically or not.

In the following subsections, we present the results of our analysis of the use of alphabetical authorship. General trends and disciplinary differences are discussed in Subsections 3.1 and 3.2, the use of partial alphabetical authorship (i.e., some but not all authors of a publication are listed alphabetically) is studied in Subsection 3.3, and the relation between the use of alphabetical authorship and the number of authors of a publication is considered in Subsection 3.4. Finally, in Subsection 3.5, we briefly discuss the availability of the data underlying our analysis for follow-up studies.

### 3.1. General trends

All results presented in this paper relate to multi-author publications. Single-author publications are not considered. We start by noting that the percentage multi-author publications has increased quite substantially over time. This can be seen in Figure 2. In 1981, 66.2% of all publications had multiple authors. In 2011, the percentage multi-author publications was 89.1%. In the rest of this section, the term 'publication' always refers to a multi-author publication. Furthermore, the term 'alphabetical publication' refers to a multi-author publication with alphabetically listed authors.

Figure 3 indicates that the percentage alphabetical publications has decreased more or less linearly during the past three decades, from 32.2% in 1981 to 15.9% in 2011. Hence, in 30 years time, the percentage alphabetical publications has halved. Does this mean that the alphabetical authorship system has become less popular among scientists and has been set aside in favor of other systems, such as a system in which authors are listed based on their contribution to a publication? This is not necessarily the case. As we have discussed, a decrease in the percentage alphabetical publications may simply be caused by an increase in the average number of authors per publication. Such an increase would lower the probability that the authors of a publication incidentally end up in alphabetical order.

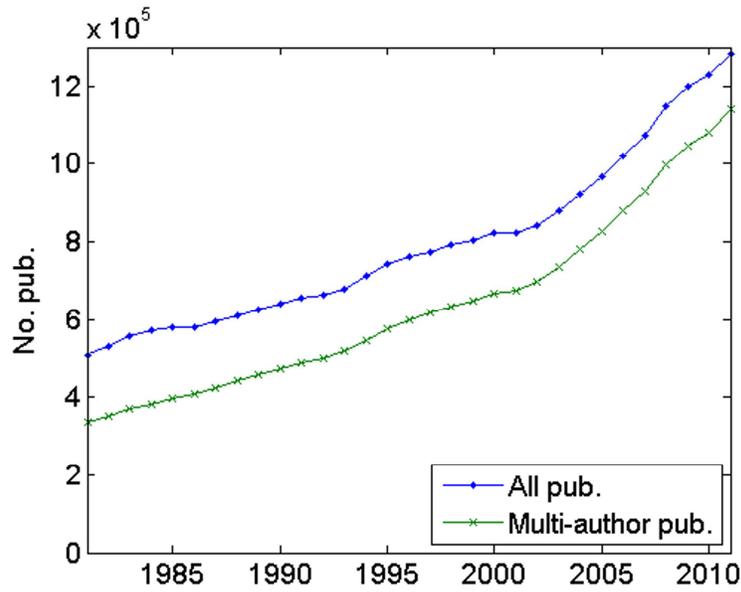

Figure 2. Trend in the total number of publications and in the number of multi-author publications.

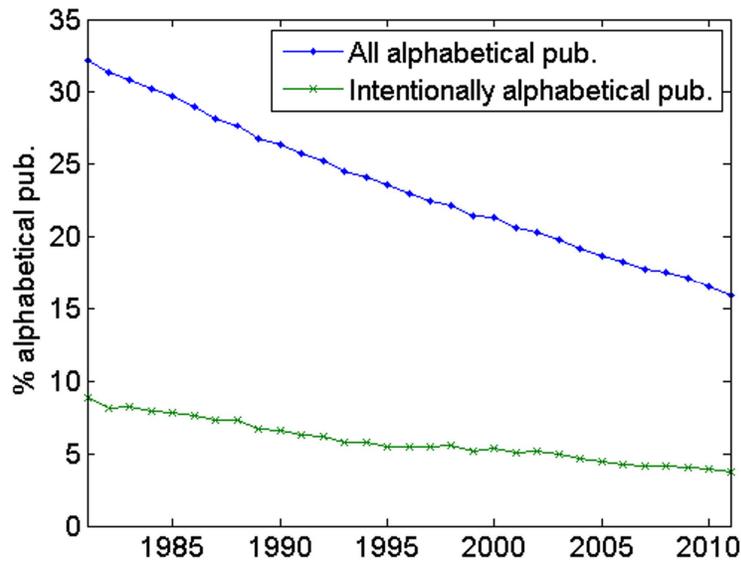

Figure 3. Trend in the percentage alphabetical publications and in the percentage intentionally alphabetical publications.

It is well known that the average number of authors per publication has increased over time (e.g., Wuchty, Jones, & Uzzi, 2007). This is confirmed by Figure 4, which shows the trend in the average number of authors per publication between 1981 and 2011. The question of course is to what extent the increase in the average number of authors per publication is responsible for the decrease in the percentage alphabetical publications. To answer this question, we need to know the percentage intentionally alphabetical publications, that is, the percentage publications in which the authors have intentionally chosen to list their names alphabetically. This percentage can be estimated using (2) and (3) in Section 2. Figure 3 reveals that the percentage intentionally alphabetical publications has decreased consistently during the past 30 years. It has declined from 8.9% in 1981 to 3.7% in 2011, indicating that the overall decrease in alphabetical authorship can only partly be explained by the increase in the average number of authors per publication. Hence, both incidental and intentional alphabetical authorship have decreased over time.

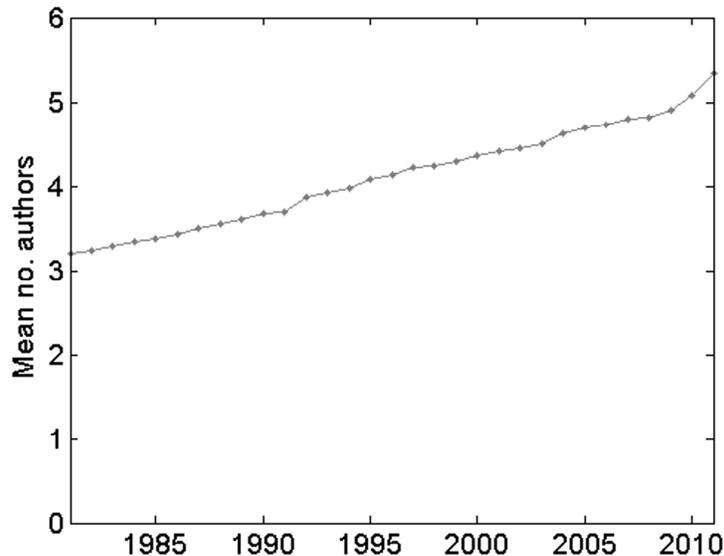

Figure 4. Trend in the average number of authors per publication (based on multi-author publications only).

Figure 3 also shows that in our period of analysis the percentage intentionally alphabetical publications has always been rather low. In other words, looking at science as a whole, it is quite uncommon for the authors of a publication to intentionally choose to list their names alphabetically. Straightforward linear extrapolation of the dashed line in Figure 3 even suggests that somewhere between 2030 and 2035 the phenomenon of intentional alphabetical authorship may have disappeared altogether. Of course, our analysis so far has completely ignored disciplinary differences in authorship practices. These differences will be analyzed in the next subsection.

**3.2. Disciplinary differences**

To analyze disciplinary differences in the use of alphabetical authorship, we rely on the Web of Science journal subject categories to define fields of science. Some publications belong to multiple subject categories. We count these publications fractionally in each of the subject categories to which they belong. Our focus is on publications from the period 2007–2011. Of the 250 subject categories, there are 27 with fewer than 1000 multi-author publications in this period. These categories, which are mostly in the arts and humanities, are not included in our analysis.

Figure 5 shows the distribution of both the percentage alphabetical publications and the percentage intentionally alphabetical publications for the 223 subject categories in our analysis. As can be seen, there are large disciplinary differences in the use of alphabetical authorship. In some subject categories, the use of alphabetical authorship is quite common. In many other subject categories, however, intentional alphabetical authorship is a virtually non-existent phenomenon. Alphabetical authorship does occur in these subject categories, but it is almost always of an incidental nature. In other words, the names of the authors of a publication may be listed alphabetically, but this has usually not been the authors' intentional choice.

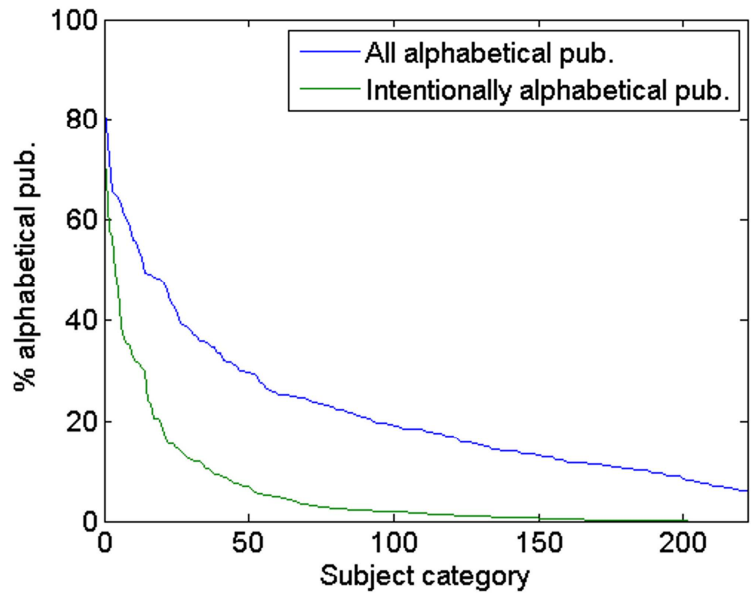

Figure 5. Distribution of the percentage alphabetical publications and of the percentage intentionally alphabetical publications for 223 subject categories.

Which are the subject categories in which intentional alphabetical authorship is a more common phenomenon? These subject categories are listed in Table 1. The table shows the 25 subject categories that have at least 15% intentionally alphabetical publications. In addition to the percentage intentionally alphabetical publications of a subject category, the table also reports the average number of authors per publication, the overall percentage alphabetical publications, and the average alphabetization score. (Average alphabetization scores will be discussed in the next subsection.) As can be seen in Table 1, subject categories with a relatively high percentage intentionally alphabetical publications can be found mostly, but not exclusively, in the social sciences and humanities and in mathematics. There turn out to be four subject categories with more than 50% intentionally alphabetical publications. These subject categories are 'Mathematics', 'Business, finance', 'Economics', and 'Physics, particles & fields'.[3] In these categories, the authors of more than half of all

---

[3] The frequent use of alphabetical authorship in economics is well documented in the literature. See for instance Efthyvoulou (2008), Einav and Yariv (2006), Engers, Gans, Grant, and King (1999), Frandsen and Nicolaisen (2010), Joseph, Laband, and Patil (2005), Laband (2002), Laband and Tollison (2000, 2006), and Van Praag and Van Praag (2008). Frandsen and Nicolaisen also report the frequent use of

publications have intentionally chosen to list their names alphabetically. We note that, in comparison with the other subject categories listed in Table 1, the subject category 'Physics, particles & fields' is a somewhat special case. Publications in this subject category have an average number of authors of 18.8, which is much larger than what is observed for the other subject categories. We will analyze the effect of the number of authors of a publication in more detail in Subsection 3.4.

Table 1. The 25 subject categories with the highest percentage intentionally alphabetical publications.

| Subject category | Mean no. authors per pub. | % alphabetical pub. | % intentionally alphabetical pub. | Mean alphabetization score (in %) |
|---|---|---|---|---|
| Mathematics | 2.4 | 83.3% | 73.3% | 73.7% |
| Business, finance | 2.6 | 78.9% | 68.3% | 68.7% |
| Economics | 2.5 | 72.3% | 58.0% | 58.6% |
| Physics, particles & fields | 18.8 | 64.4% | 56.7% | 64.1% |
| Social sciences, mathematical methods | 2.6 | 65.2% | 49.5% | 49.8% |
| Mathematics, applied | 2.6 | 63.4% | 46.2% | 46.6% |
| Philosophy | 2.2 | 65.6% | 38.8% | 39.3% |
| Political science | 2.4 | 60.8% | 36.3% | 36.9% |
| Statistics & probability | 2.7 | 55.8% | 35.2% | 35.4% |
| International relations | 2.5 | 59.9% | 35.0% | 35.2% |
| Computer science, theory & methods | 3.2 | 48.6% | 32.6% | 33.6% |
| Physics, mathematical | 3.0 | 49.6% | 31.8% | 32.4% |
| Law | 2.6 | 55.5% | 31.5% | 31.7% |
| Industrial relations & labor | 2.7 | 52.5% | 30.4% | 32.1% |
| History | 2.4 | 58.7% | 29.9% | 30.3% |
| Planning & development | 2.7 | 49.1% | 24.1% | 25.3% |
| Operations research & management science | 2.8 | 45.1% | 23.5% | 24.1% |
| Area studies | 2.3 | 53.9% | 20.6% | 21.3% |
| Urban studies | 2.7 | 46.6% | 20.5% | 21.7% |
| Public administration | 2.6 | 48.1% | 20.3% | 20.8% |
| History & philosophy of science | 2.6 | 47.6% | 18.5% | 17.8% |
| Mathematics, interdisciplinary applications | 2.9 | 39.2% | 16.5% | 16.6% |
| Language & linguistics theory | 2.5 | 48.2% | 15.6% | 16.3% |
| Demography | 2.8 | 42.1% | 15.4% | 15.5% |
| Humanities, multidisciplinary | 2.6 | 49.0% | 15.4% | 16.4% |

Let's now look somewhat closer at the subject categories 'Mathematics', 'Business, finance', 'Economics', and 'Physics, particles & fields'. For each of these subject categories, Figure 6 shows the trend in the percentage intentionally alphabetical publications between 1981 and 2011. As can be seen, the subject category 'Business, finance' experienced a large increase in the percentage intentionally alphabetical publications during the 1990s. The other three subject categories display a relatively stable pattern over the past three decades. In recent

---

alphabetical authorship in high energy physics (which in our analysis is represented by the subject category 'Physics, particles & fields').

years, however, the percentage intentionally alphabetical publications has been clearly decreasing in the subject categories 'Mathematics' and 'Economics'. This development is in line with the general trend shown in Figure 3.

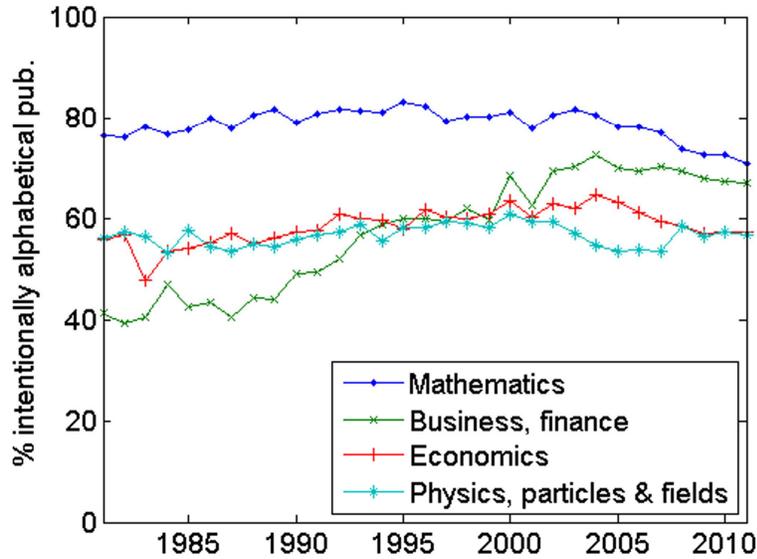

Figure 6. Trend in the percentage intentionally alphabetical publications in four subject categories.

**3.3. Partial alphabetical authorship**

Until now, we have assumed that the authors of a publication either choose their names to be listed alphabetically or not. There has been no room for the situation in which a combination of alphabetical and non-alphabetical criteria is used to determine a publication's authorship order. An example of such a situation could be a publication with five authors where there is one author who has clearly made the largest contribution while the other four authors have all made smaller contributions, each of them of about the same size. In this situation, the name of the author with the largest contribution may be listed first, while the names of the other authors may be listed next in alphabetical order. This type of authorship could be referred to as partial alphabetical authorship.

To measure not only full but also partial alphabetical authorship, we introduce the alphabetization score of a publication. The alphabetization score of publication $i$ is given by

$$s_i = 2\frac{m_i}{n_i - 1} - 1, \qquad (4)$$

where $n_i$ denotes the number of authors of publication $i$ and $m_i$ denotes the number of pairs of consecutive author names that are listed in alphabetical order. For instance, in the case of a publication authored by Smith, Johnson, Jones, and Williams, $m_i$ would be equal to two. This is because we have two pairs of alphabetically listed author names, namely the pair Johnson and Jones and the pair Jones and Williams. The alphabetization score of the publication would consequently be equal to $2 \times 2 / (4 - 1) - 1 = 0.33$. Alphabetization scores range between –1 and 1. A score of –1 indicates that there are no pairs of alphabetically listed author names, while a score of 1 indicates full alphabetical authorship. If a publication's authorship order is determined by a non-alphabetical criterion, one would expect on average half of the pairs of consecutive author names to be listed in alphabetical order. This corresponds with an alphabetization score of 0. We note that alphabetization scores cannot be calculated for single-author publications.

For a given set of publications, we can calculate both the estimated proportion intentionally alphabetical publications (as discussed in Section 2) and the average alphabetization score. It is important to see the relation between these two numbers. If for each publication in our set the authorship order is determined either by an alphabetical or by a non-alphabetical criterion, but not by a combination of these two, then the estimated proportion intentionally alphabetical publications and the average alphabetization score will be approximately equal.[4] Differences between the two numbers will arise if for some publications in our set the authorship order is determined by a combination of alphabetical and non-alphabetical criteria. This partial alphabetical authorship will increase the average alphabetization score of the publications in our set. We note that in the case of publications with two authors there can be no partial alphabetical authorship. Because of this, if $n_i = 2$, (3) in Section 2 will always yield the same result as (4) above.

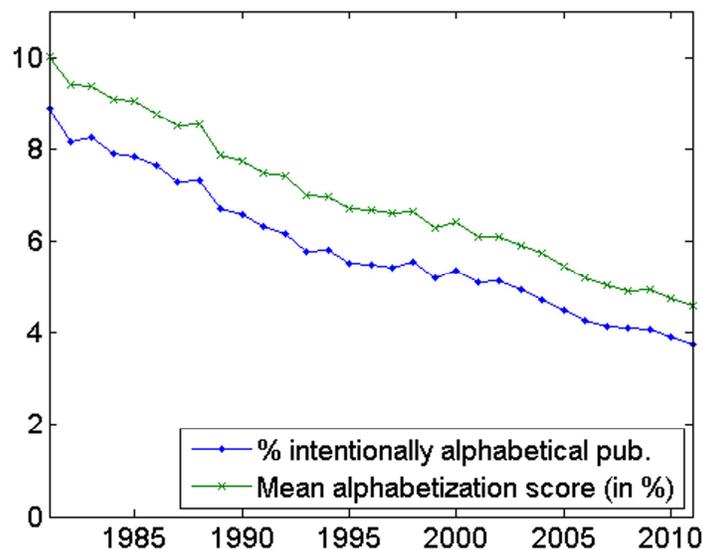

Figure 7. Trend in the percentage intentionally alphabetical publications and in the average alphabetization score.

---

[4] To see this, notice that the average alphabetization score of the publications for which the authorship order is determined by an alphabetical criterion (i.e., the intentionally alphabetical publications) will be 1, while the average alphabetization score of the remaining publications will be approximately 0. The overall average alphabetization score will therefore be approximately equal to the proportion intentionally alphabetical publications.

For science as a whole, Figure 7 shows the trend in the percentage intentionally alphabetical publications and in the average alphabetization score between 1981 and 2011. To facilitate comparison, the average alphabetization score is expressed as a percentage of the maximum score of 1. The average alphabetization score turns out to be consistently higher than the percentage intentionally alphabetical publications. This is a clear indication of the effect of partial alphabetical authorship. Notice, however, that the effect is not very large.

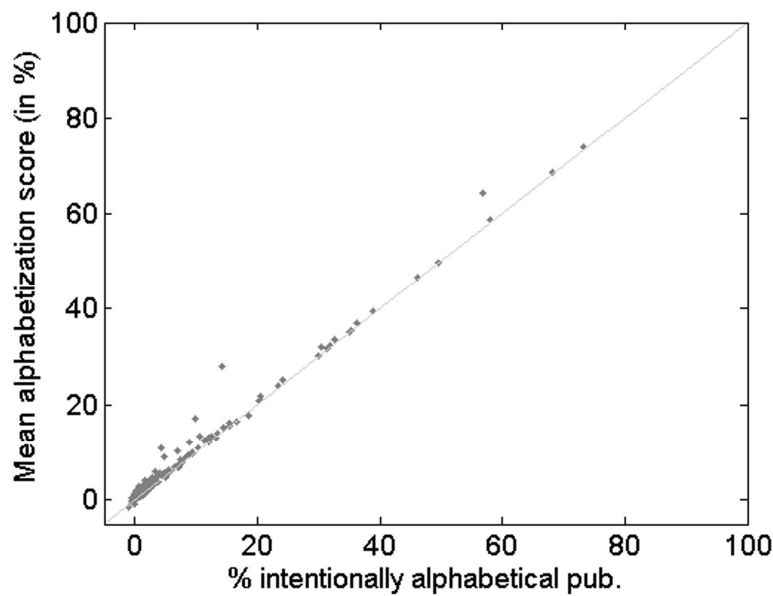

Figure 8. Scatter plot of the percentage intentionally alphabetical publications and the average alphabetization score for 223 subject categories.

Table 2. The five subject categories with the largest difference between the average alphabetization score and the percentage intentionally alphabetical publications.

| Subject category | Mean no. authors per pub. | % alphabetical pub. | % intentionally alphabetical pub. | Mean alphabetization score (in %) |
|---|---|---|---|---|
| Physics, nuclear | 9.9 | 26.5% | 14.3% | 28.1% |
| Physics, particles & fields | 18.8 | 64.4% | 56.7% | 64.1% |
| Astronomy & astrophysics | 8.9 | 23.3% | 10.0% | 17.0% |
| Nuclear science & technology | 5.7 | 15.7% | 4.3% | 11.0% |
| Instruments & instrumentation | 8.8 | 18.6% | 4.8% | 9.1% |

Let's now look at possible disciplinary differences. Like in the previous subsection, our analysis relies on publications from the period 2007–2011. Only the 223 subject categories with at least 1000 multi-author publications in this period are considered. Figure 8 reveals that in most subject categories hardly any effect of partial alphabetical authorship can be found. However, there are a small number of subject categories in which partial alphabetical authorship turns out to have a quite significant effect. These subject categories have an average alphabetization score that is substantially higher than their percentage intentionally alphabetical publications. The five subject categories for which the difference is largest are listed in Table 2.

Although these five subject categories are all in the natural sciences, the overall picture emerging from Tables 1 and 2 indicates that most subject categories with a high average alphabetization score can be found in the social sciences and humanities and in mathematics. This is similar to what we observed earlier for the percentage intentionally alphabetical publications. We further note that the five subject categories listed in Table 2 all have a relatively large average number of authors per publication. This is something we will analyze in more detail in the next subsection.

**3.4. Relation between alphabetical authorship and the number of authors of a publication**

Does there exist a relation between the use of alphabetical authorship and the number of authors of a publication? For the period 2007–2011, Figure 9 shows for our 223 subject categories how the average alphabetization score relates to the average number of authors per publication. A clear U-shape can be observed. Subject categories that have either a small or a large average number of authors per publication tend to have a relatively high average alphabetization score. As can be seen in Tables 1 and 2, these subject categories can be found in the social sciences and humanities, in mathematics, and in physics. Subject categories whose average number of authors per publication is in between the extremes usually have a very low average alphabetization score. Many of these subject categories are in the medical and life sciences. We note that a picture very similar to Figure 9 emerges when looking at the percentage intentionally alphabetical publications instead of the average alphabetization score (not shown). The main difference is that for subject categories with a large average number of authors per publication the percentage intentionally alphabetical publications is somewhat lower than the average alphabetization score (see also Table 2).

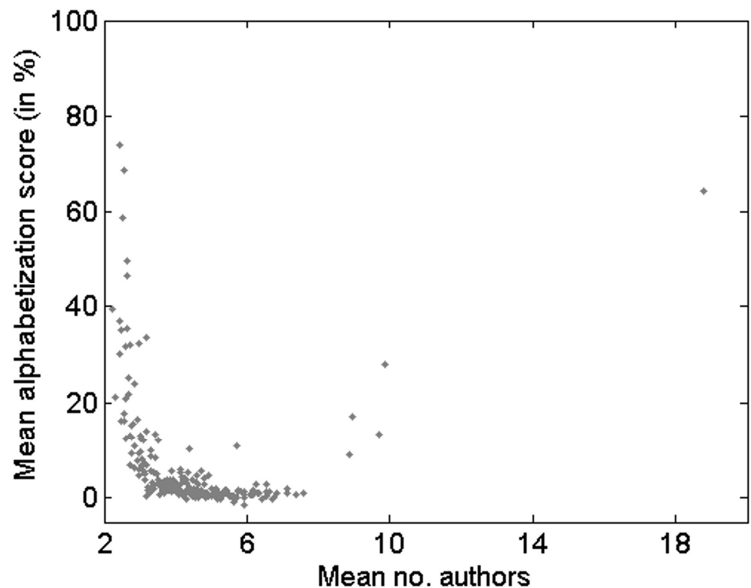

Figure 9. Scatter plot of the average number of authors per publication and the average alphabetization score for 223 subject categories.

We now consider the relation between the use of alphabetical authorship and the number of authors of a publication not at the subject category level but at the level of individual publications. Our analysis is based on publications from the period 2007–2011 in all fields of science. As can be seen in Figure 10, the percentage alphabetical publications is quite high for publications with only two or three authors. For a large part, this is of course caused by incidental alphabetical authorship. However, looking at the percentage intentionally alphabetical publications, we still observe relatively high scores for publications with two or three authors. This seems to be due to the fact that the use of alphabetical authorship is more common in fields with a small average number of authors per publication, especially in fields in the social sciences and humanities and in mathematics (see Table 1). For publications with only a small number of authors, the average alphabetization score more or less coincides with the percentage intentionally alphabetical publications. As discussed in the previous subsection, this is because with only a small number of authors there is no or almost no room for partial alphabetical authorship.

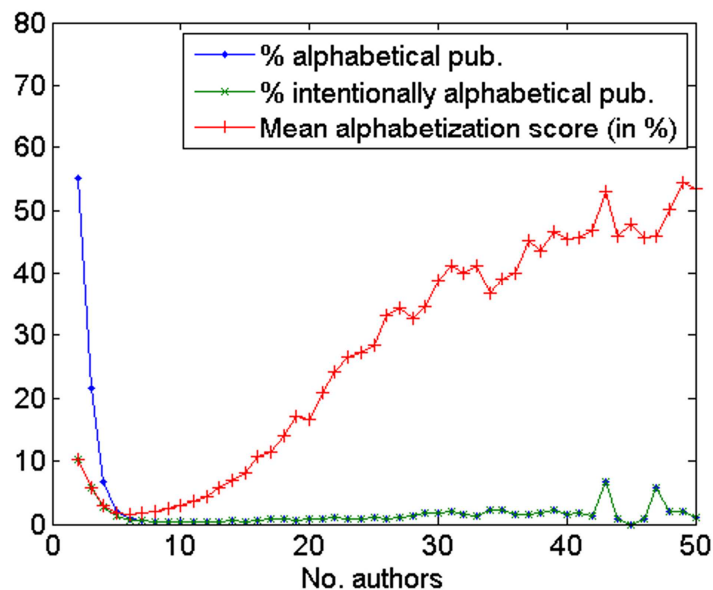

Figure 10. Relation between the number of authors of a publication and the percentage alphabetical publications, the percentage intentionally alphabetical publications, and the average alphabetization score.

For publications with larger numbers of authors, the percentage alphabetical publications and the percentage intentionally alphabetical publications cannot be distinguished anymore in Figure 10. The reason for this is that with larger numbers of authors incidental alphabetical authorship is highly unlikely to occur (see Section 2). For publications with more than five authors, Figure 10 indicates that the percentage alphabetical publications is close to zero, although the percentage appears to slightly increase with the number of authors of a publication. Interestingly, the average alphabetization score shows a very different picture. It increases rapidly with the number of authors of a publication, reaching a score above 50% for publications with 50 authors. Figure 10 does not provide statistics for publications with more than 50 authors. There turn out to be 4072 of these 'hyperauthorship' (Cronin, 2001) publications in the period 2007–2011 (0.08% of the total number of publications). Among the 4072 publications, there are only 2.7% alphabetical publications, even though the average alphabetization score equals 77.0%. This seems to indicate that partial alphabetical authorship is frequently used in the case of publications with many authors, while full alphabetical authorship is not. However, a more detailed analysis reveals that this conclusion is not correct. In a random sample of 30 non-alphabetical publications with more than 50 authors, we found that in 12 publications (40%) author names actually do seem to be listed alphabetically.[5] In these publications, the rules for determining the alphabetical order of author names seem to be slightly different from the rules that we use in this paper (as discussed in the beginning of Section 3). For instance, prefixes in last names (e.g., 'DE', 'DI', or 'VAN') are sometimes treated differently. Another problem is that in some rare cases the last name of an author is not registered correctly in the Web of Science database. This may for instance happen with Spanish authors who have two last names.

### 3.5. Availability of data for follow-up analyses

Many additional analyses are possible based on the data collected for the research presented in this paper. We have therefore made the data freely available at www.ludowaltman.nl/alphabetical_authorship/. The data are provided at the level of subject category-publication year combinations. Anyone interested in the phenomenon of alphabetical authorship is invited to use the data for follow-up analyses.

## 4. Conclusions

In this paper, we have studied the use of alphabetical authorship in scientific publishing. Special attention has been paid to the distinction between intentional and incidental alphabetical authorship. The main findings of our analysis can be summarized as follows:

- During the past three decades, there has been a consistently declining trend in the use of alphabetical authorship. In 1981, the authors of 8.9% of all publications in the Web of Science database intentionally chose to list their names alphabetically. This has decreased to 3.7% in 2011.
- The use of alphabetical authorship is most common in the social sciences and humanities and in mathematics. There are four Web of Science subject categories with more than 50% intentionally alphabetical publications in the period 2007–2011: 'Mathematics', 'Business, finance', 'Economics', and 'Physics, particles & fields'.
- The use of partial alphabetical authorship (i.e., some but not all authors of a publication are listed alphabetically) is most common in natural science fields with a relatively large average number of authors per publication.
- The use of alphabetical authorship is relatively more common in fields that have either a small or a large average number of authors per publication. A similar conclusion can be drawn at the level of individual publications rather than fields.

---

[5] Based on our random sample, 'alphabetical hyperauthorship' seems to be much more common in physics than in biomedical research. This is in line with Birnholtz (2006), who reports that the standard practice in high energy physics is to list the names of the authors of a hyperauthorship publication in alphabetical order.

As mentioned in Section 3.5, the data underlying our analysis are freely available and can be used for follow-up studies.

Our findings may be helpful to identify proper credit assignment strategies for multi-author publications. In particular, in fields in which there is a substantial use of alphabetical authorship, giving more credits to the first author of a publication than to the other authors is clearly not a good strategy. On the other hand, this strategy may be appropriate in fields in which alphabetical authorship is a virtually non-existent phenomenon. Nevertheless, even in such fields, assuming that the first author of a publication is the most significant contributor may not always be warranted. In some fields, the last author may for instance play an important role as well (e.g., Shapiro, Wenger, & Shapiro, 1994). Furthermore, in many publications, the first author and the corresponding author are different, which suggests that the corresponding author may also have made an important contribution.[6] Proper credit assignment seems even more difficult in the case of 'hyperauthorship' publications. From a credit assignment point of view, these publications, with tens or even hundreds of authors, many of whom have probably made only a very indirect contribution (Birnholtz, 2006), may well require to be handled in a completely different way than ordinary publications.

Finally, it is important to be aware of the limitations of the analysis presented in this paper. There are three important limitations that need to be mentioned. First, all results that we have reported are dependent on what is covered by the Web of Science database and what is not. In addition, the coverage of the database has changed over time and includes more and more journals. This changing database coverage may have affected the results of our trend analyses. Second, all results reported in this paper depend on the exact way in which alphabetical authorship is defined (as discussed in the beginning of Section 3). As we have seen in Subsection 3.4, because of the use of slightly different rules for determining the alphabetical order of author names, the actual use of alphabetical authorship may be higher than what we have reported, in particular in the case of publications with many authors. Third, our estimation of the proportion intentionally alphabetical publications relies on some assumptions, and these assumptions may not hold exactly in practice. In fact, the phenomenon of partial alphabetical authorship analyzed in Subsection 3.3 already contradicts the assumptions that we make and may cause the proportion intentionally alphabetical publications to be somewhat overestimated.

## Appendix

In this appendix, we prove that $\hat{p}$ in (2) is an unbiased estimator of $\bar{p}$ in (1).

We first consider the probability that the names of the authors of publication $i$ are listed alphabetically. This probability equals

$$\Pr(a_i = 1) = p_i + (1 - p_i)\frac{1}{n_i!}. \tag{5}$$

---

[6] Analyzing the difference between first authors and corresponding authors (referred to as reprint authors in the Web of Science database), it turns out that in recent years more than one-third of all publications (including single-author publications) had a corresponding author who is different from the first author. In addition, there turns out to be a clear increasing trend in the proportion publications with a corresponding author who is not the first author. There are also publications that have multiple corresponding authors (Hu, 2009), but the Web of Science database does not seem to register this.

To see this, recall that $p_i$ equals the probability that the authors of publication $i$ intentionally choose to list their names alphabetically. Consequently, $1 - p_i$ equals the probability that the authors of publication $i$ choose to list their names based on a non-alphabetical criterion. When a non-alphabetical criterion is used, there is a probability of $1 / n_i!$ that incidentally the names of the authors are listed alphabetically. This follows from the assumption that in the case of a non-alphabetical criterion all possible orderings of author names are equally likely to be observed.

The expected value of $\hat{p}_i$ in (3) is given by

$$\mathrm{E}(\hat{p}_i) = \Pr(a_i = 0) \frac{-\frac{1}{n_i!}}{1 - \frac{1}{n_i!}} + \Pr(a_i = 1). \tag{6}$$

Setting $\Pr(a_i = 0) = 1 - \Pr(a_i = 1)$, substituting (5), and simplifying yields $\mathrm{E}(\hat{p}_i) = p_i$. It now follows that the expected value of $\hat{p}$ in (2) equals

$$\mathrm{E}(\hat{p}) = \frac{1}{N} \sum_{i=1}^{N} \mathrm{E}(\hat{p}_i) = \frac{1}{N} \sum_{i=1}^{N} p_i = \bar{p}. \tag{7}$$

Hence, the expected value of $\hat{p}$ equals $\bar{p}$ in (1). This proves that $\hat{p}$ is an unbiased estimator of $\bar{p}$.

## References

Abbas, A.M. (2011). Weighted indices for evaluating the quality of research with multiple authorship. *Scientometrics*, *88*(1), 107–131.

Birnholtz, J.P. (2006). What does it mean to be an author? The intersection of credit, contribution, and collaboration in science. *Journal of the American Society for Information Science and Technology*, *57*(13), 1758–1770.

Cronin, B. (2001). Hyperauthorship: A postmodern perversion or evidence of a structural shift in scholarly communication practices? *Journal of the American Society for Information Science and Technology*, *52*(7), 558–569.

Efthyvoulou, G. (2008). Alphabet economics: The link between names and reputation. *Journal of Socio-Economics*, *37*(3), 1266–1285.

Egghe, L., Rousseau, R., & Van Hooydonk, G. (2000). Methods for accrediting publications to authors or countries: Consequences for evaluation studies. *Journal of the American Society for Information Science*, *51*(2), 145–157.

Einav, L., & Yariv, L. (2006). What's in a surname? The effects of surname initials on academic success. *Journal of Economic Perspectives*, *20*(1), 175–188.

Engers, M., Gans, J.S., Grant, S., & King, S.P. (1999). First-author conditions. *Journal of Political Economy*, *107*(4), 859–883.

Frandsen, T.F., & Nicolaisen, J. (2010). What is in a name? Credit assignment practices in different disciplines. *Journal of Informetrics*, *4*(4), 608–617.

Galam, S. (2011). Tailor based allocations for multiple authorship: A fractional *gh*-index. *Scientometrics*, *89*(1), 365–379.

Hagen, N.T. (2008). Harmonic allocation of authorship credit: Source-level correction of bibliometric bias assures accurate publication and citation analysis. *PLoS ONE*, *3*(12), e4021.

Hagen, N.T. (2010). Harmonic publication and citation counting: sharing authorship credit equitably – not equally, geometrically or arithmetically. *Scientometrics*, *84*(3), 785–793.

Hodge, S.E., & Greenberg, D.A. (1981). Publication credit. *Science*, *213*, 950.

Hu, X. (2009). Loads of special authorship functions: Linear growth in the percentage of "equal first authors" and corresponding authors. *Journal of the American Society for Information Science and Technology*, *60*(11), 2378–2381.

Hu, X., Rousseau, R., & Chen, J. (2010). In those fields where multiple authorship is the rule, the *h*-index should be supplemented by role-based *h*-indices. *Journal of Information Science*, *36*(1), 73–85.

Joseph, K., Laband, D.N., & Patil, V. (2005). Author order and research quality. *Southern Economic Journal*, *71*(3), 545–555.

Laband, D., & Tollison, R. (2006). Alphabetized coauthorship. *Applied Economics*, *38*(14), 1649–1653.

Laband, D.N. (2002). Contribution, attribution and the allocation of intellectual property rights: Economics versus agricultural economics. *Labour Economics*, *9*(1), 125–131.

Laband, D.N., & Tollison, R.D. (2000). Intellectual collaboration. *Journal of Political Economy*, *108*(3), 632–662.

Liu, X.Z., & Fang, H. (2012a). Fairly sharing the credit of multi-authored papers and its application in the modification of *h*-index and *g*-index. *Scientometrics*, *91*(1), 37–49.

Liu, X.Z., & Fang, H. (2012b). Modifying *h*-index by allocating credit of multi-authored papers whose author names rank based on contribution. *Journal of Informetrics*, *6*(4), 557–565.

Marušić, A., Bošnjak, L., & Jerončić, A. (2011). A systematic review of research on the meaning, ethics and practices of authorship across scholarly disciplines. *PLoS ONE*, *6*(9), e23477.

Sekercioglu, C.H. (2008). Quantifying coauthor contributions. *Science*, *322*, 371.

Shapiro, D.W., Wenger, N.S., & Shapiro, M.F. (1994). The contributions of authors to multiauthored biomedical research papers. *Journal of the American Medical Association*, *271*(6), 438–442.

Van Praag, C.M., & Van Praag, B.M.S. (2008). The benefits of being economics professor A (rather than Z). *Economica*, *75*(300), 782–796.

Wuchty, S., Jones, B.F., & Uzzi, B. (2007). The increasing dominance of teams in production of knowledge. *Science*, *316*, 1036–1039.